\documentstyle [prb,aps,epsf]{revtex}
\begin{document}

\draft

\title{ Effect of disorder on quantum 
phase transitions in anisotropic\\
 XY spin chains in a transverse field}

\author{J. E. Bunder and Ross H. McKenzie\cite{email}}

\address{School of Physics, University of New
South Wales, Sydney 2052, Australia}

\date{December 31, 1998}
\maketitle

\begin{abstract}
We present some exact results for the effect
of disorder on the critical properties
of an anisotropic XY spin chain in a transverse
field. The continuum limit of the corresponding
fermion model is taken and in various cases
results in a Dirac equation with a random mass.
Exact analytic techniques can then be used to
evaluate the density of states and the localization
length. In the presence of disorder the ferromagnetic-paramagnetic
or Ising transition of the model is in the same universality
class as the random transverse field Ising model 
solved by Fisher using a real space renormalization
group decimation technique (RSRGDT). If there is only randomness
in the anisotropy of the magnetic exchange
then the anisotropy transition
(from a ferromagnet in the $x$ direction to a ferromagnet in the
 $y$ direction)
is also in this universality class.
However, if there is randomness in the isotropic part of
the exchange or in the transverse field then
in a non-zero transverse field the anisotropy transition
is destroyed by the disorder.
We show that in the Griffiths' phase near the Ising
transition that the ground state energy has an
essential singularity. 
The results obtained for the dynamical critical exponent,
typical correlation length,
and for the temperature dependence of the specific heat
near the Ising transition agree with the
results of the RSRGDT and numerical work.\\
\\
\end{abstract}


\section{Introduction}

An important feature of many low-dimensional models
of strongly interacting electrons is that
they exhibit quantum phase transitions, i.e.,
they undergo a phase transition at zero temperature
as some parameter is varied.\cite{sondhi}
Experimental realization of this occurs in
heavy fermion materials such as CeCu$_{6-x}$Au$_x$ 
which undergo an antiferromagnetic-paramagnetic phase
transition induced by pressure.\cite{itp,nature}
Near the critical point unconventional
metallic behaviour is observed and is enhanced by
the presence of disorder.\cite{itp,castro}
It has also been proposed that a quantum critical point
plays an important role in cuprate superconductors.\cite{cuprate}
Quantum phase transitions in the presence of impurities or
disorder also occur in
$^4$He and $^3$He absorbed in porous media,~\cite{today2}
 superconductor-insulator transitions in dirty thin films,~\cite{herbut}
the delocalization transition in the quantum Hall effect
and the metal-insulator transition in doped semiconductors.~\cite{dob1}

Compared to thermal
phase transitions in disorder-free systems these transitions are poorly
understood because many of the theoretical methods (e.g., exact solutions,
the renormalization group and $\epsilon$ expansions) that have proven
so useful for
pure systems at non-zero temperatures\cite{chaikin} are difficult to
implement for disordered systems.~\cite{today1}  
These phase transitions are associated with particularly rich physics
such as 
large differences between average and typical (i.e., most probable) 
behaviour, new universality classes,
logarithmic scaling and ``Griffiths phases'',~\cite{griffiths} in which
susceptibilities diverge  even though
there are only short-range correlations.
Low energy properties of the system are dominated
by extremely rare configurations of the system.
It has recently been proposed that
Griffiths phases can lead to unconventional
metallic behaviour.\cite{castro,dob}
                                                                     
Fisher recently made an exhaustive study of the
effect of randomness on what is arguably the simplest
model to undergo a quantum phase transition:
the transverse field Ising spin chain.~\cite{dsf}  He used
a real space renormalization group decimation technique
(RSRGDT), originally developed by Dasgupta and Ma,~\cite{ma} which he claimed is exact near the critical point.
Fisher  found the phase diagram
(which included a Griffiths phase near the critical point), all of the
critical exponents (some of which are irrational, as shown in Table \ref{table0})
 and scaling functions for the
magnetization and correlation functions in an external field.
It is striking that
the latter have never been derived for the disorder-free case
but can be derived in the presence of disorder because
distributions become extremely broad near the critical point.
Many of Fisher's results have been confirmed by numerical work.~\cite{dsf4,yr,young}
The RSRGDT has now also been used to study
the effect of disorder on dimerized~\cite{hyman} and
anisotropic spin-${1 \over 2}$ chains,
spin-1 chains,~\cite{yang,monthus}
chains with random spin sizes,~\cite{west}
quantum Potts and clock chains,~\cite{senthil}
and diffusion in a random environment.~\cite{dsf3,le} Possible experimental realizations of random spin chains
are given in Table \ref{table0}. There is a direct connection between the critical behaviour of the random transverse field Ising chain and random
 walks in a disordered environment.~\cite{dsf3,le,rieger,igloi,igloi2}
 Fisher's results have also been related to the Kondo lattice in one dimension.~\cite{hg}

An important question is whether some of the same interesting
physics occurs in higher dimensional models.
Senthil and Sachdev did find this to be the case in a
dilute quantum Ising system near a percolation transition.\cite{senthil2}
It is particularly interesting that some of the most
striking features that Fisher found in the one-dimensional model
(a variable dynamical critical exponent which 
diverges at the critical point and the average and typical
correlations are associated with different critical exponents)
have recently been found in the two-dimensional 
random transverse field Ising model.\cite{rieger1}

The outline of the paper is as follows.
In Section II we introduce the model,
an anisotropic XY spin chain in a transverse
field, where all the exchange integrals and 
transverse field are random.
A similar model was also recently studied numerically
using the density-matrix renormalization group.\cite{sweden}
A Jordan-Wigner transformation is then used to
map the model onto a non-interacting fermion model.
Section III contains a brief summary of the
known properties of the disorder-free model
that are needed to understand the rest of the paper.
In Section IV we take  the
 continuum limit of the fermion model for  various cases.
The Ising transition and the anisotropy transition with
only randomness in the anisotropy
that  results in a Dirac equation with a random mass.
The isotropic XX chain in a transverse field with
randomness in the exchange and/or transverse field
reduces to a Dirac equation with a random complex mass.
Mapping the spin chain to these Dirac equations
has the advantage that a number of different exact
analytic techniques can then be used to
evaluate the density of states and the localization
length. The properties of the solutions corresponding
to the universality class of the random transverse field
Ising model are then discussed in Section V.
By examining the energy dependence of the density
of states we evaluate the dynamical critical exponent,
show the existence of a  Griffiths' phase near the transition, and show
that the ground state energy has an
essential singularity at the transition. 
We also present results for thermodynamic properties
and the typical correlation length.
The results obtained for the dynamical critical exponent,
typical correlation length,
and for the temperature dependence of the specific heat
near the Ising transition agree with the
results of the RSRGDT and numerical work.
Since our approach is explicitly exact our results
are consistent with Fisher's claim that the 
RSRGDT gives exact results for critical properties.
In Section VI we point out that the properties
of the incommensurate solution are such that
it implies that the anisotropy transition is destroyed
in a non-zero transverse field if there is randomness
in the isotropic exchange or in the transverse field.
A brief report of some of the results presented here
appeared previously.\cite{mck0}

\section{The model}
\label{sec-model}
The Hamiltonian to be considered is that of an anisotropic
XY spin chain in a transverse field:
\begin{equation}
H = -\sum_{n=1}^L \left(J_n^x \sigma^x_n \sigma^x_{n+1} +
 J_n^y \sigma^y_n \sigma^y_{n+1} +
 h_n \sigma^z_n \right).
\label{ham}
\end{equation}
The ${\sigma^a_n}$, $(a=x,y,z)$, are Pauli spin matrices. This is a quantum model because the Pauli
matrices do not commute with one another. The interactions,
 $J_n^x$,  $J_n^y$, 
and transverse fields, $h_n$, are independent
random variables 
with Gaussian distributions.
All the results given in this paper are for this
ferromagnetic case but also hold for the
antiferromagnetic case. By means of a spin rotation $J_n^x$ and $h_n$ can always be
chosen to be non negative. We shall assume that $J_n^y$ is also non negative so that there is no frustration in the system. 
The  average values will be denoted
\begin{equation}
\langle J_n^x \rangle \equiv J^x, \ \ \ \ \langle J_n^y \rangle \equiv J^y, \ \ \ \
\langle h_n \rangle \equiv h .
\label{avge}
\end{equation}
The deviation of the random variables from their average values is assumed to be small, relative to the average value. We can write our three parameters in terms of a random part and an averaged part,
\begin{equation}
J_n^{a}=J^a+\delta J_n^a \ \ (a=x,y),\qquad h_n=h+\delta h_n.
\end{equation}
The random variables are uncorrelated between sites and
have variances
\begin{equation}
\langle (J_n^a - J^a)^2 \rangle  \equiv (\delta J^a)^2 \ \ (a=x,y), \qquad
\langle (h_n-h)^2 \rangle \equiv \delta h^2.
\label{sigma}
\end{equation}

For $J_n^y=0$ the model is the random transverse field
Ising spin chain which is the quantum analog of the
two-dimensional Ising model with random coupling
in one direction, introduced by McCoy and Wu,~\cite{mw}
and studied by Shankar and Murthy.~\cite{sm}

At zero temperature and in
the  absence of disorder the model undergoes two
distinct quantum phase transitions.~\cite{dennijs}
Both transitions are second order.
The phase diagram is shown in Fig. \ref{phased}.
The transition  at
$J^x + J^y= h $ from a paramagnetic to a ferromagnetic phase
will be referred to as the {\it Ising transition}.~\cite{pfeuty}
The  transition at
$ J^x=J^y $ for $ h < (J^x + J^y) $ from a
ferromagnet with magnetization in the $x$ direction to one with
magnetization in the $y$ direction will be
referred to as the {\it anisotropic transition}.~\cite{dsf2,lsm}
This paper considers the effect of disorder on
these transitions.

\subsection{Mapping to a fermion model}
We perform a Jordan-Wigner transformation which maps the Pauli spin matrices
in (\ref{ham})
 onto spinless fermions.~\cite{lsm,katsura}
The Pauli spin matrices
\begin{equation}
\sigma^x =
 \left(\matrix{0 & 1 \cr 1 & 0} \right)
\ \ \ \ \
\sigma^y = 
 \left(\matrix{0 & -i \cr i & 0} \right)
\ \ \ \ \
\sigma^z = 
 \left(\matrix{1 & 0 \cr 0 & -1} \right)
\ \ \ \ \
\label{pauli}
\end{equation}
satisfy the algebra
\begin{equation}
[\sigma^a,\sigma^b]=
2i \epsilon_{abc}\sigma^c\qquad  (\sigma^a)^2=1.
\end{equation}
Define the following new operators on each site,
\begin{equation}
a_n^{\dag}=\frac{1}{2}(\sigma_n^x+i\sigma_n^y)\qquad a_n=\frac{1}{2}(\sigma_n^x-i\sigma_n^y).
\end{equation}
The inverse transformation is
\begin{equation}
\sigma_n^x=a_n^{\dag}+a_n\qquad \sigma_n^y=
i(a_n-a_n^{\dag})\qquad \sigma_n^z=1-2a_na_n^{\dag}.
\end{equation}
Using these definitions the following relations can be obtained,
\begin{equation}
\{a_n^{\dag},a_n\}=1,\qquad a_n^2
=a_n^{\dag 2}=0,\qquad [a_m^{\dag},a_n]
=[a_m^{\dag},a_n^{\dag}]=[a_m,a_n]=0\qquad m\neq n
\end{equation}
which show that the $a_n$'s and $a^{\dag}_n$'s
are neither fermion operators nor boson operators.
 The Hamiltonian, (\ref{ham}), in terms of these new operators is,
\begin{equation}
H=-\sum_{n=1}^L[(J_n^x+J_n^y)
(a_n^{\dag}a_{n+1}+a_ia_{n+1}^{\dag})
+(J_n^x-J_n^y)(a_n^{\dag}a_{n+1}^{\dag}+a_na_{n+1})+h_n(1-2a_na_n^{\dag})].
\label{eq:Ha}
\end{equation}

Now consider a second transformation,
\begin{eqnarray}
c_n&=&\exp(\pi i\sum_{j=1}^{n-1}a_j^{\dag}a_j)a_i\nonumber\\
c_n^{\dag}&=&a_n^{\dag}\exp(-\pi i\sum_{j=1}^{n-1}a_j^{\dag}a_j).
\end{eqnarray}
The $c_n$'s and $c^{\dag}_n$'s are fermion operators
satisfying the following anti-commutation relations,
\begin{equation}
\{c_m,c_n^{\dag}\}=\delta_{mn}\qquad \{c_m,c_n\}=\{c_m^{\dag},c_n^{\dag}\}=0.
\end{equation}
These fermions can be viewed as kinks or domain walls in the
local magnetization.~\cite{dennijs}
 The Hamiltonian is now,
\begin{equation}
H=-\sum_{n=1}^L[(J_n^x+J_n^y)(c_n^{\dag}c_{n+1}-c_nc_{n+1}^{\dag})
+(J_n^x-J_n^y)(c_n^{\dag}c_{n+1}^{\dag}-c_nc_{n+1})+h_n(c_n^{\dag}c_n-c_nc_n^{\dag})].
\label{eq:Hc}
\end{equation}
The boundary terms have been neglected since they do not contribute to the thermodynamic limit.

\section{Solution of the disorder-free case}

\label{sec-nodis}

The model in the absence of disorder
has been solved previously.~\cite{barouch,sachdev,dennijs}
We now highlight certain aspects of the solution
that will turn out to be 
particularly relevant to the effect of disorder.
In the disorder free case $J_n^x=J^x$, $J_n^y=J^y$ and $h_n=h$ so that
\begin{equation}
H=-\sum_{n=1}^L[(J^x+J^y)(c_n^{\dag}c_{n+1}-c_nc_{n+1}^{\dag})+(J^x-J^y)(c_{n}^{\dag}c_{n+1}^{\dag}-c_nc_{n+1})+h(c_n^{\dag}c_n-c_nc_n^{\dag})].\label{eq:Hdf}
\end{equation}
The case $h=0$ corresponds to the anisotropic $XY$ spin chain and was first solved by Lieb, Schultz and Mattis.~\cite{lsm} The case $J^y=0$ is the transverse field Ising chain and was first solved by Pfeuty.~\cite{pfeuty}

We introduce the Fourier transform of the fermion operators
\begin{eqnarray}
c_n&=&\frac{1}{\sqrt{L}}\sum_kc_ke^{ink}\nonumber\\
c_n^{\dag}&=&\frac{1}{\sqrt{L}}\sum_kc_k^{\dag}e^{-ink}.
\label{ft}
\end{eqnarray}
Periodic boundary conditions ($c_n=c_{n+L}$) require the wave vector, $k$, to take the following discrete values,
\begin{equation}
k=\frac{2\pi m}{L},\qquad m=-\frac{1}{2}L,\ldots,0,1,\ldots,\frac{1}{2}L-1,
\end{equation}
assuming $L$ to be even. 
Substituting (\ref{ft}) in the Hamiltonian (\ref{eq:Hdf}) gives
\begin{equation}
H=-\sum_k[2((J^x+J^y)\cos k+h)c_k^{\dag}c_k+i(J^x-J^y)\sin k(c_k^{\dag}c_{-k}^{\dag}+c_kc_{-k})-h].\label{eq:Hd}
\end{equation}
This Hamiltonian maybe diagonalised by the Bogoluibov transformation,
\begin{eqnarray}
c_k^{\dag}=\cos\phi(k)b_k^{\dag}+i\sin\phi(k)b_{-k}\nonumber\\
c_k=\cos\phi(k)b_k-i\sin\phi(k)b_{-k}^{\dag},
\end{eqnarray}
where the $b_k$'s and $b_k^{\dag}$'s are operators with fermion statistics, and
\begin{equation}
\tan(2\phi(k))=\frac{(J^x-J^y)\sin k}{(J^x+J^y)\cos k+h}.
\end{equation}
This gives,
\begin{equation}
H=-\sum_kE(k)[b_k^{\dag}b_k-1/2]\label{eq:HE}
\end{equation}
where
\begin{equation}
E(k)=2(h^2+(J^x-J^y)^2+2h(J^x+J^y)\cos k+4J^xJ^y\cos^2k)^{1/2},
\label{edisp}
\end{equation}
and we have used,
$\sum_k\cos k=0$.

\begin{figure}
\centerline{\epsfxsize=8.0cm \epsfbox{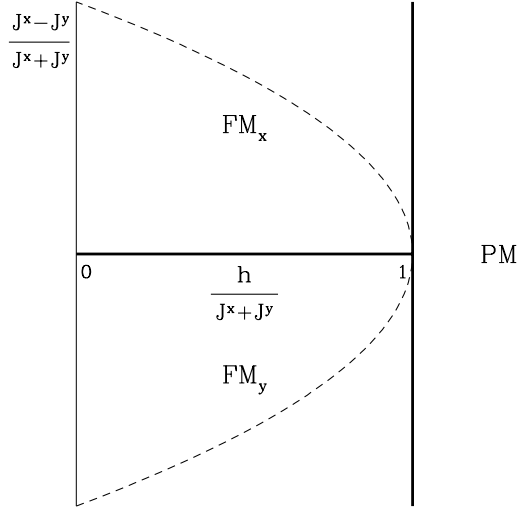}}
 \caption{
Phase diagram for the anisotropic XY spin chain
in a transverse field at zero temperature
and in the absence of disorder.
The heavy lines represent second order phase
transitions. The horizontal line will be
referred to as the anisotropic transition
and the vertical line as the Ising transition.
PM denotes a paramagnetic phase and FM$_x$
 denotes an Ising ferromagnet with magnetization in the $x$ direction.
To the right of the dashed line 
the energy gap in the excitation spectrum always occurs at
the Brillouin zone boundary (compare Fig. \protect\ref{figdisp}). 
To the left of the dashed line gap occurs at 
a wave vector that is incommensurate with the lattice.
\label{phased}} \end{figure}

\begin{figure}
\centerline{\epsfxsize=8.0cm \epsfbox{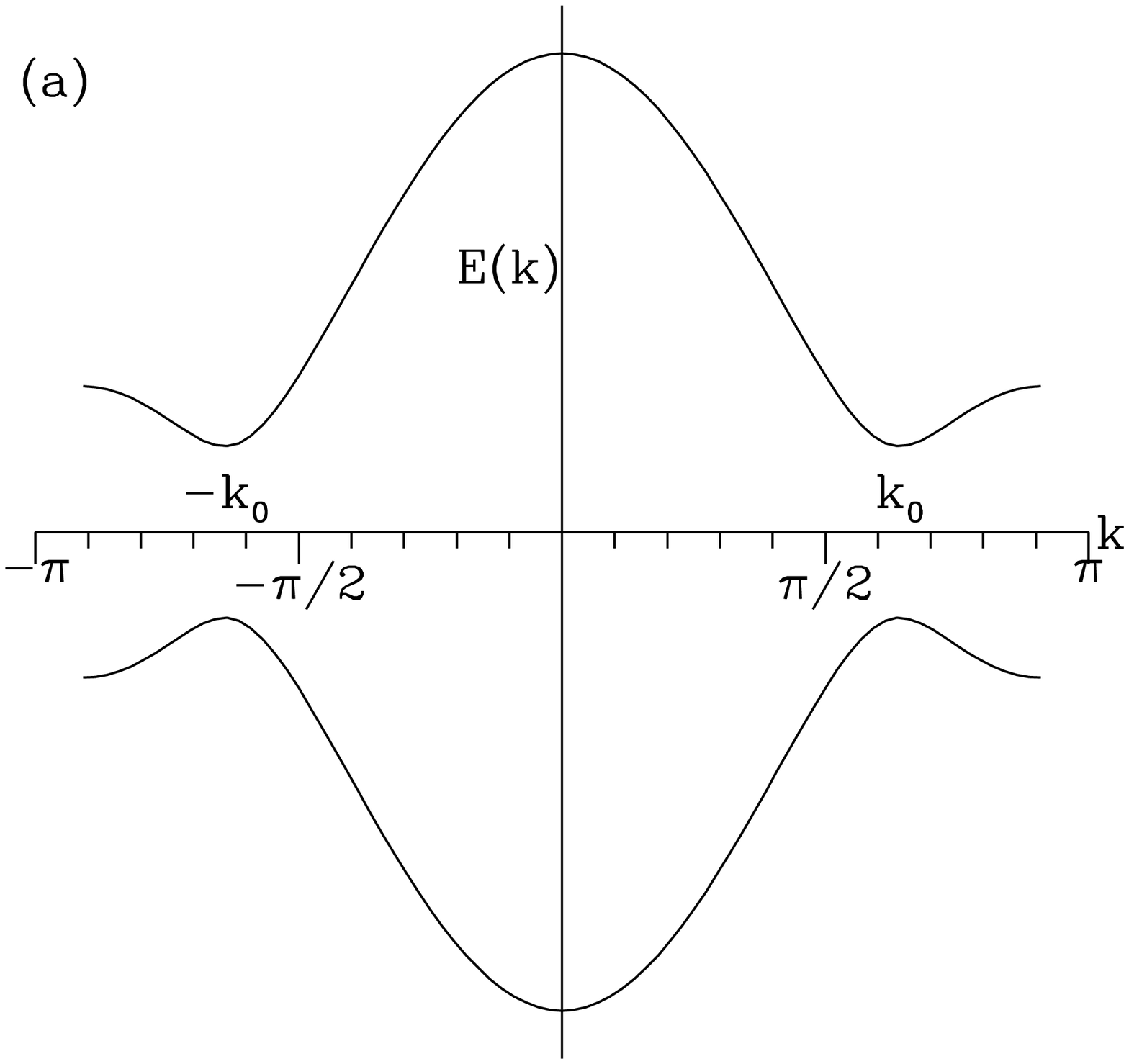}}
\centerline{\epsfxsize=8.0cm \epsfbox{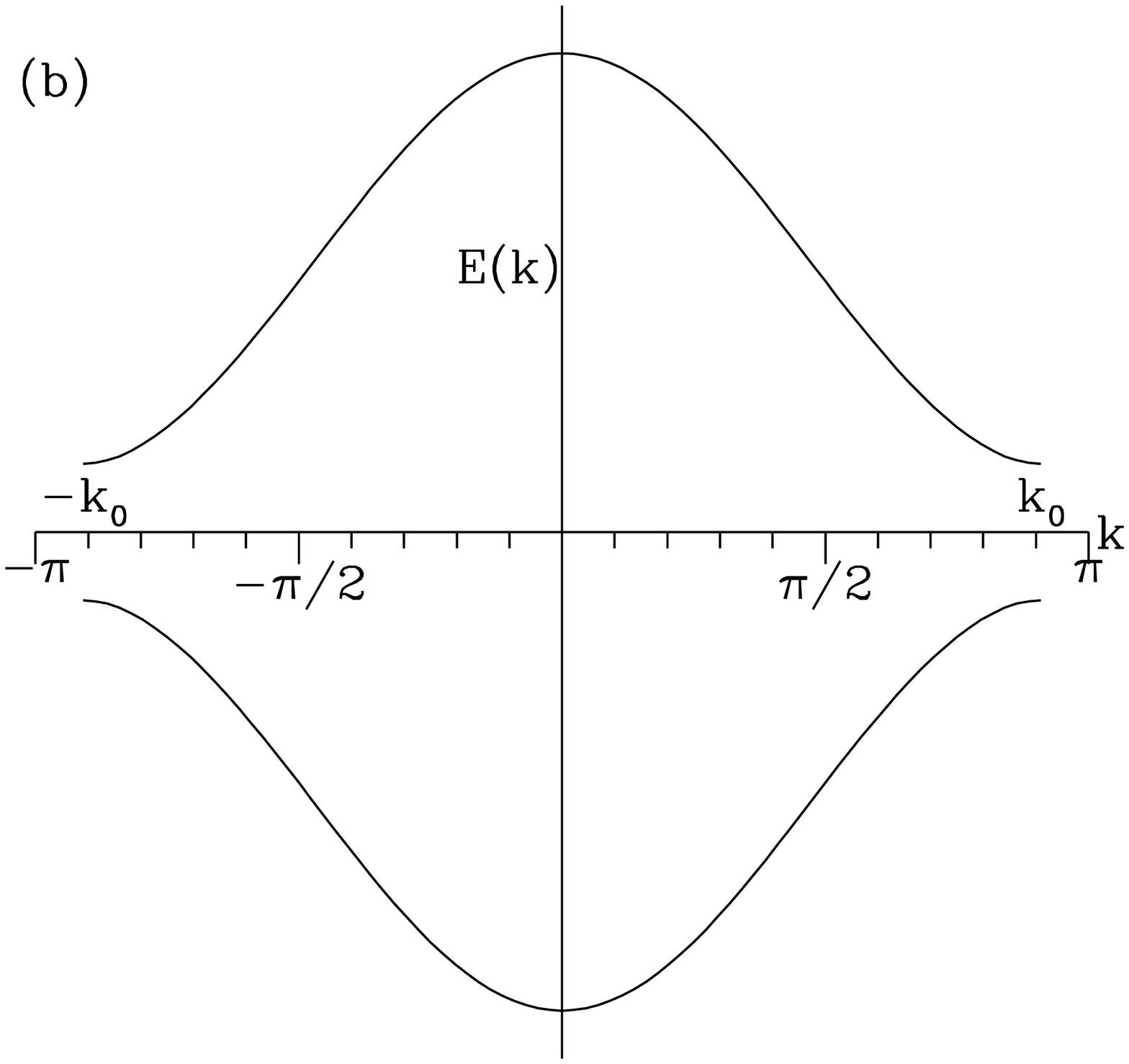}}
 \caption{
Typical dispersion relations for the excitation
spectrum of the Hamiltonian (\protect\ref{ham}) in the absence of disorder.
The two cases shown correspond to when the
quantity $\alpha$, defined in equation (\protect\ref{alpha}), is
(a) larger than one and (b) less than one.
Note that for (a) the energy gap always occurs
at the Brillouin zone boundary whereas for (b)
it occurs at a wave vector that is incommensurate 
with the reciprocal lattice vectors.
The cases (a) and (b) occur in regions
of the phase diagram  to the right and
left of the dashed line in Fig. \protect\ref{phased}.
(The commensurate case also occurs on the vertical line $h=0$:
then $k_0 =\pi/2$).
\label{figdisp}} \end{figure}

The energy gap will occur at a wave vector $k_0$ such that
\begin{equation}
 {dE(k=k_0) \over dk} =0.
\end{equation}
We shall always take $k_0$ to be the positive solution of the above equation. Because the energy is symmetric in $k$ there will be two energy gaps at $\pm k_0$. To describe the solutions of this equation it is convenient to define
\begin{equation}
\alpha=-\frac{h(J^x+J^y)}{4J^xJ^y},
\label{alpha}
\end{equation}
which is always negative for non negative $J^x$, $J^y$ and $h$.
The energy gap wave vector, $k_0$, is given by
\begin{eqnarray}
\cos k_0=\alpha \qquad  \alpha&>&-1,\nonumber\\
 k_0=\pi \qquad \alpha&<&-1.
\end{eqnarray}
Typical dispersion curves for these two cases
are shown in Fig. \ref{figdisp}.
If we define, 
\begin{equation}
\gamma \equiv 
\frac{J^x-J^y}{J^x+J^y},
\end{equation}
and express $\alpha$ as
\begin{equation}
\alpha=-\frac{h}{(J^x+J^y)(1-\gamma^2)}.  \end{equation}
then, the boundary between the two cases may be defined by
\begin{equation}
\frac{h}{J^x+J^y}=1-\gamma^2.
\end{equation}
This boundary is shown as a dashed line in Fig. \ref{phased}.
The two cases correspond to a commensurate
($\alpha < -1$) and an incommensurate 
($\alpha > -1$) phase.~\cite{dennijs}
It will turn out that the effect of
disorder on these two phases is very different.


The energy gap at $k=k_0$ is $2\Delta$ where
\begin{equation}
\Delta=E(k_0).
\end{equation}
The system is at criticality when the gap vanishes, $\Delta=0$. When $\alpha<-1$ ($k_0=\pi$) the gap vanishes along the line $h=J^x+J^y$.
 We shall refer to the corresponding phase transition as the
{\it  Ising transition}.
 When $\alpha>-1$ the energy gap vanishes along the line $J^x=J^y$,
 providing $h<J^x+J^y$. We shall refer to this transition as the 
{\it anisotropic transition}.
The lines along which the gap vanishes are shown as solid lines in
Fig. \ref{phased}.

The critical behaviour is determined by those low energy states near the energy gap where $k\sim k_0$. If $k-k_0$ is small the energy can be written as a Taylor series, 
\begin{equation}
E(k)^2=\Delta^2+v_0^2(k-k_0)^2+\ldots
\label{smallk}
\end{equation}
where
\begin{eqnarray}
v_0&=&2[4J^xJ^y-h(J^x+J^y)\cos k_0-8J^xJ^y\cos^2 k_0]^{1/2}\nonumber\\
\Delta&=&2[h^2+(J^x-J^y)^2+2h(J^x+J^y)\cos k_0+4J^xJ^y\cos^2 k_0]^{1/2}.
\label{vel0}
\end{eqnarray}

\subsection{The ground state energy}
\label{gse0}

The ground state energy, $\epsilon$, of the Hamiltonian (\ref{eq:Hdf}) is
the energy of the filled Fermi sea 
\begin{equation}
\epsilon(\Delta)=-\int_{-\pi}^{\pi}\frac{dk}{2\pi}E(k)
\end{equation}
where $E(k)$ is given by (\ref{edisp}).
To discover the nature of the singularity at $\Delta=0$
 we differentiate the above integral with respect to $\Delta^2$. For small $\Delta$ the differentiated integral is dominated by those low energies states close to the energy gap. Hence, we need only consider those states determined by the low energy dispersion
relation (\ref{smallk}),
\begin{eqnarray}
\frac{\partial\epsilon}{\partial\Delta^2}
&=&\int_{k_0-k_c}^{k_0+k_c}\frac{dk}{4\pi}\frac{1}
{\sqrt{\Delta^2+v_0^2(k-k_0)^2}}\\
&=&
-\frac{1}{2\pi v_0}\ln\left[\frac{\Delta}{2v_0 k_c}\right]
\end{eqnarray}
where $k_c$ is a cut-off wave vector.
 Integrating with respect to $\Delta^2$
gives 
\begin{equation}
\epsilon(\Delta)-\epsilon(0)=
\frac{\Delta^2}{4\pi v_0}
\left(1- 2\ln\left[\frac{\Delta}{2v_0 k_c}\right]\right).  \end{equation}
The singularity of the ground state energy is thus logarithmic.
The critical exponent $\alpha$, defined by
$\epsilon(\Delta) \sim \Delta^{2 - \alpha}$,
is $\alpha=0^+$. This critical exponent corresponds to the specific heat 
critical exponent of the corresponding two-dimensional classical Ising model.

\subsection{The magnetization and correlation length}

Barouch and McCoy~\cite{barouch}
calculated the magnetization and correlation
functions for the disorder-free model.
Its {\it et al.} considered the case $\delta=0$.~\cite{korepin}
Further analysis was done by Damle and Sachdev.~\cite{sachdev}
The magnetization, $M^x\equiv \langle \sigma^x_n \rangle$,
 and the correlation length, $\xi$,
are defined by the asymptotic behavior ($r \to \infty$) of the correlation function,
\begin{equation}
\langle \sigma^x_n \sigma^x_{n+r} \rangle
 \to (M^x)^2  + { A  \over r^2} 
\exp( - r/ \xi)
 \end{equation}
where $A$ is a constant. If $h>J^x+J^y$ the system is a paramagnet and the magnetization is zero. 
If $h < J^x + J^y$ and $J^x>J^y$ the system is a ferromagnet in the $x$ direction, and the magnetization is,
\begin{equation}
(M^x)^2 =
(-1)^r { 2 \gamma^{1 \over 2} \over 1 + \gamma} 
\left[1 - \left({h \over  J^x + J^y}\right)^2  \right]^{1 \over 4}.
 \end{equation}
This implies that the critical exponent $\beta$ is $1/8$ for the Ising transition (approaching the transition as a ferromagnet)
and $1/4$ for the anisotropic transition.  

The correlation length, $\xi$, is given by
\begin{equation}
\exp\left({-1 \over \xi}\right) \sim |\lambda_2|^{-2}
\end{equation}
where,
\begin{equation}
\lambda_2=\frac{\frac{h}{J^x+J^y}-[(\frac{h}{J^x+J^y})^2-(1-\gamma^2)]^{\frac{1}{2}}}{1-\gamma}.
\end{equation}
This quantity is real (complex) in the commensurate (incommensurate)
phase. As a result
\begin{equation}
\exp\left({-1 \over \xi}\right) \sim\left\{\begin{array}{ll}
\lambda_2^{-2}\qquad &\mbox{commensurate}\\ 
\frac{1-\gamma}{1+\gamma} &\mbox{incommensurate}.
\end{array}\right.
\end{equation}
This implies that the critical exponent
$\nu = 1$ for both the Ising and anisotropy transitions.

On the Ising critical line
\begin{equation}
\langle \sigma^x_n \sigma^x_{n+r} \rangle
\sim { 1 \over r^{1/4}}
 \end{equation} and the critical exponent $\eta = 5/4$.
On the anisotropic critical line
\begin{equation}
\langle \sigma^x_n \sigma^x_{n+r} \rangle
\sim { 1 \over r^{1/2}}
 \end{equation}
and the critical exponent $\eta = 3/2$. The anisotropic transition has the same critical
behavior as a pair of decoupled Ising models.
The critical exponents for the Ising transition
are summarized in Table \ref{tableo}.

\section{The Continuum Limit}
\label{sec-cont}
We shall now look at the effect of disorder on the critical behaviour. To do this we take the continuum limit of the disordered Hamiltonian written in terms of Fermi operators, (\ref{eq:Hc}), when the system is near criticality. We will assume that, for weak disorder, the phase transitions of the disordered system are close to the phase transitions of the disorder free system. Those fermion states most effected by the addition of disorder will be those low energy states near the energy gap, that is, those with wave vectors near $\pm k_0$. 

The Hamiltonian may be broken into slowly and rapidly varying parts. This is done by replacing the Fermi operator, $c_n$, with two slowly varying functions, $\psi_R(n)$ and $\psi_L(n)$, which describe right and left movers respectively,
\begin{equation}
c_n=\frac{1}{\sqrt{2}}[e^{-ik_0n}\psi_R(n)+e^{ik_0n}\psi_L(n)].\label{eq:slow}
\end{equation}
The exponential terms represent the rapidly varying part of $c_n$. From the anticommutation relations of the Fermi operators, $c_n$ and $c_n^{\dag}$, it is possible to derive anticommutation relations for the slowly varying functions,
\begin{equation}
\{\psi_R^{\dag}(n),\psi_R(m)\}=\{\psi_L^{\dag}(n),\psi_L(m)\}=\delta_{nm}\label{eq:anticom},
\end{equation}
and all other combinations are zero.

\subsection{Ising Transition}
When $k_0=\pm\pi$, equation (\ref{eq:slow}) may be simplified to,
\begin{equation}
c_n=(-1)^n\psi(n),
\end{equation}
with,
\begin{equation}
\{\psi^{\dag}(n),\psi(m)\}=\delta_{nm}.\label{eq:slow2}
\end{equation}
After substituting this into the Hamiltonian, (\ref{eq:Hc}), we take the continuum limit. To do this we transform the discrete variable, $n$, into a continuous variable, $x$, and we write,
\begin{equation}
\psi(n)=\psi(x=n),\qquad \psi(n+1)=\psi(x)+\partial_x\psi(x).
\end{equation}
The function $\psi$ is slowly varying and so its derivative is very small. Where appropriate we can neglect these derivative terms. The disorder is assumed to be small, hence terms combining both derivative terms and disordered terms may be neglected.
Substituting these approximations into the Hamiltonian and replacing the sum over $n$ with an integral over $x$ gives,
\begin{equation}
H=\sum_{n=1}^L(\psi^{\dag},\psi)[iJ^-\sigma^y\partial_x+(J^+(x)-h(x))\sigma^z]\left(\begin{array}{c}
\psi\\
\psi^{\dag}\end{array}\right),
\end{equation}
where
\begin{equation}
J^{\pm}(x)=J^x_{n=x}\pm J^y_{n=x},\qquad h(x)=h_{n=x}.
\end{equation}
By performing the following rotation into a new set of Pauli spin matrices, with $k_0=\pm\pi$,
\begin{eqnarray}
\sigma^y&=&\frac{\Delta J^+\sin k_0}{2J^-(J^++h\cos k_0)}\sigma^1+\frac{v_0(h+J^+\cos k_0)}{2J^-(J^++h\cos k_0)}\sigma^3\nonumber\\
\sigma^z&=&-\frac{\Delta\cos k_0}{2(J^++h\cos k_0)}\sigma^1+\frac{v_0\sin k_0}{2(J^++h\cos k_0)}\sigma^3\nonumber\\
\sigma^x&=&\sigma^2,\label{eq:rotation}
\end{eqnarray}
where $\Delta$ and $v_0$ are defined equation (\ref{vel0}), the Hamiltonian becomes,
\begin{equation}
H=\frac{1}{2}\int dx\Psi(x)^{\dag}[-iv_0\sigma^3\partial_x+V(x)\sigma^2]\Psi(x),\label{eq:H1}
\end{equation}
where,
\begin{equation}
V(x)=2|J^+-h|\pm 2(\delta J^+(x)-\delta h(x)),
\end{equation} 
and $\delta J^+(x)$ and $\delta h(x)$ are the random parts and $J^+$ and $h$ are the average parts of $J^+(x)$ and $h(x)$ respectively.
The function $V(x)$ is real and its average value is $\Delta$, the energy gap of the pure system. For the case of no disorder (\ref{eq:H1}) was derived by Shankar.~\cite{dirac} The case of the transverse field Ising chain
with randomness only in $J^x$ or $h$ was derived by Balents and Fisher.~\cite{balents}  
The fact that the Ising transition is described by the same equation for any
anisotropy shows that it will be in the same universality class as the
random transverse field Ising chain studied by Fisher.

\subsection{Anisotropic Transition}
Near the anisotropic transition we must use the more general decomposition of the Fermi operators shown in equation (\ref{eq:slow}). As was done near the Ising transition we replace the discrete variable, $n$, with a continuous variable, $x$, and replace discrete differences with derivatives. Next, we remove those disordered terms which are negligible. We shall make one approximation which was not necessary near the Ising transition. We will neglect all rapidly varying terms. A rapidly varying term may be neglected because its integral will vanish. Terms involving the product of two rapidly terms may not be neglected since the two rapid variations may produce a slowly varying part. The exponential terms are rapidly varying and the random terms may have slowly and rapidly varying parts. Combining all these approximations gives the following Hamiltonian,
\begin{eqnarray}
H=-\int dx&&\left[\frac{1}{2}(J^+(x)\cos k_0+h(x))(\psi_R^{\dag}\psi_R+\psi_L^{\dag}\psi_L)+iJ^-(x)\sin k_0\psi_L^{\dag}\psi_R^{\dag}-\frac{h(x)}{2}\right.\nonumber\\
&&+\frac{J^+}{2}e^{-ik_0}\psi_R^{\dag}\partial_x\psi_R+\frac{J^+}{2}e^{ik_0}\psi_L^{\dag}\partial_x\psi_L+\frac{J^-}{2}e^{-ik_0}\psi_R^{\dag}\partial_x\psi_L^{\dag}+\frac{J^-}{2}e^{ik_0}\psi_L^{\dag}\partial_x\psi_R^{\dag}\nonumber\\
&&\left.+e^{2ik_0x}(\delta J^+(x)e^{ik_0}+\delta h(x))\psi_R^{\dag}\psi_L\right]+\mbox{h.c.}\label{eq:Hsum}
\end{eqnarray}
When $k_0=-\frac{\pi}{2}$, $J^x(x)=J^y(x)=\frac{t_x}{2}$ and $h(x)=0$, this Hamiltonian is equivalent to that obtained by Balents and Fisher.~\cite{balents}
The Hamiltonian in equation (\ref{eq:Hsum}) can be simplified in two particular cases. 

\subsubsection{Case I ($\delta J_n^x = - \delta J_n^y$, \ \ $\delta h_n=0$)}
If the only randomness is in the anisotropy, which must be equal but opposite ($\delta J^+(x)=\delta h(x)=0$), the Hamiltonian reduces to,
\begin{eqnarray}
H=-\int dx&&(\psi_R^{\dag}, \psi_L)[(J^+\cos k_0+h)\sigma^z-J^-(x)\sin k_0\sigma^y\nonumber\\
&&+i(J^+\sin k_0\sigma^z+J^-\cos k_0\sigma^y)\partial_x]\left(\begin{array}{c}
\psi_R\\
\psi_L^{\dag}\end{array}\right).
\end{eqnarray}
On performing the rotation in equation (\ref{eq:rotation}), the Hamiltonian can be simplified to,
\begin{equation}
H=\frac{1}{2}\int dx(\psi_R^{\dag},\psi_L)[-iv_0\sigma^3\partial_x+V(x)\sigma^1]\left(\begin{array}{c}
\psi_R\\
\psi_L^{\dag}\end{array}\right)\label{eq:H2}.
\end{equation}
We have used the definition of $k_0$ on the anisotropic critical line, $\cos k_0=-\frac{h}{J^+}$, and defined, 
\begin{equation}
V(x)=2|J^-|\sin k_0\pm 2\delta J^-(x)\sin k_0.
\end{equation}
Note that, like the Ising transition, $V$ is real and its average value is $\Delta$.
Hence, in this case the anisotropy transition is in the same universality
class as the random transverse field Ising chain.

\subsubsection{Case II ($J_n^x = J_n^y$)}
\label{sec-caseII}
Another special case of the Hamiltonian in equation (\ref{eq:Hsum}) is when the is no anisotropy ($J^-(x)=0$). Note that, in the absence of disorder this restricts the model to the anisotropic critical line. It shall also be assumed that any disorder is rapidly varying. The new Hamiltonian is,
\begin{equation}
H=-\frac{1}{2}\int(\psi_R^{\dag}, \psi_L^{\dag})[-2iJ^+\sin k_0\sigma^z\partial_x+\xi(x)^*\sigma^++\xi(x)\sigma^-]\left(\begin{array}{c}
\psi_R\\
\psi_L\end{array}\right),
\end{equation} 
where $\xi(x)=2e^{-2ik_0x}(\delta J^+(x)e^{-ik_0}+\delta h(x))$ and $\sigma^{\pm}=\frac{1}{2}(\sigma^x\pm i\sigma^y)$. Since $k_0$ is incommensurate with the lattice $\xi(x)$ is complex. We have neglected a term involving the sum over the magnetic field since it is a constant. Consider the following rotation,
\begin{eqnarray}
\sigma^3&=&-\sigma^z\nonumber\\
\sigma^1&=&\sigma^x\nonumber\\
\sigma^2&=&-\sigma^y.
\end{eqnarray}
With this rotation,
\begin{equation}
H=\frac{1}{2}\int\Psi(x)^{\dag}[-iv_0\sigma^3\partial_x+\xi(x)\sigma^++\xi(x)^*\sigma^-]\Psi(x),\label{eq:H3}
\end{equation}
where $\sigma^{\pm}=\frac{1}{2}(\sigma^1\pm i\sigma^2)$. The complex function, $\xi(x)=2e^{-2ik_0x}(\delta J(x)^+e^{-ik_0}+\delta h(x))$, has the following properties,
\begin{eqnarray}
\langle\xi(x)\rangle&=&0\nonumber\\
\langle\xi(x)\xi(x')\rangle&=&0\nonumber\\
\langle\xi(x)\xi(x')^*\rangle&=&\gamma\delta(x-x'),
\end{eqnarray}
where $\gamma=4((\delta J^+)^2+(\delta h)^2)$. 

In summary, all three Hamiltonians, (\ref{eq:H1}), (\ref{eq:H2}) and (\ref{eq:H3}), can be written in the form,~\cite{mck}
\begin{equation}
H=\frac{1}{2}\int dx\Psi(x)^{\dag}[-iv_0\sigma^3\partial_x+V(x)\sigma^++V(x)^*\sigma^-]\Psi(x).\label{hamel}
\end{equation}
Note that the structure of the spinor, $\Psi$, is quite different in all three cases. The function $V(x)$ satisfies,
\begin{eqnarray}
\langle V(x)\rangle&=&\Delta\nonumber\\
\langle V(x)V(x')^*\rangle&=&\Delta^2+\gamma\delta(x-x').
\end{eqnarray}
The anisotropic case II ($J^-=0$) has $\Delta=0$. The fact that $V(x)$ is complex for the anisotropic case II will lead to qualitively different behaviour.
In fact, in that case the disorder removes the  phase transition.
 We refer to the case where $V(x)$ is real as the commensurate case (that is, the Ising transition and the anisotropic case I). The case where $V(x)$ is complex is the incommensurate case.  
The case of real $V(x)$ also describes dimerized XX spin
chains\cite{mck,fabrizio,steiner} and spin ladders.\cite{gogolin2,shelton}
The case of complex $V(x)$ also describes an XX spin chain in
a transverse field with a modulation of the exchange 
with wavevector $2k_0$.

\section{Exact solutions}
\label{sec-exact}
It is useful to define  an energy $D$
and a dimensionless parameter $\delta$
which are measures of the disorder strength
and the deviation from criticality, respectively 
\begin{equation}
D \equiv {\gamma \over v_0} \ \ \ \ \ \
\delta \equiv {\Delta \over D}.       
\label{Delta}
\end{equation}
 Note that for the Ising transition
with $J^y=0$, to leading order
in $\Delta/J^x $, for a Gaussian distribution
this parameter $\delta$
agrees with the $\delta$
defined by Fisher~\cite{dsf} and Young and Rieger,~\cite{yr}
\begin{equation}
\delta  \equiv { \langle \ln h\rangle - \langle \ln J^x\rangle \over
\langle(\ln h)^2\rangle- \langle \ln h\rangle^2 + \langle(\ln J^x)^2\rangle - \langle \ln J^x\rangle^2}.
\label{delta}
\end{equation}

The advantage of casting the problem in the form of the Hamiltonian
 (\ref{hamel})
is that the latter
has been studied extensively previously, and
{\it exact} analytic expressions given for the energy dependence of the
disorder-averaged 
density of states $ \langle\rho (E)\rangle$ and the localization length
$\lambda(E)$.
 The exact results have been found by   
Fokker-Planck equations,~\cite{ov,halp} supersymmetry,~\cite{hayn,john,balents}
 the replica trick,~\cite{bouch2,bouch,bocquet}
S-matrix summation,~\cite{golub} and the Dyson-Schmidt method.~\cite{mertsching}

Due to the one-dimensionality all the states are localized
by the disorder.
 The localization length can be found because in one dimension
it is related to the real part of the one-fermion  
 Green's function.~\cite{thou,hayn}

The density of states  and the localization length
  are related to the one-electron Green's function $G(x,x,E)$
and
can be written in terms of $f_{\delta} ^\prime (u),$
the derivative of a dimensionless function $f_{\delta}(u)$,
\begin{equation}
\hbox{Tr}\;G(E,x,x)=
 {d \over dE} {1 \over \lambda(E)} + i \pi \rho (E)
        = \pi \rho_0 f_{\delta} ^\prime ( E / D) 
\label{gf}
\end{equation}
where $\rho_0 \equiv 1/(\pi v_0)$ is the value of the density of
states at  high energies ($|E| \gg \Delta, D )$. The function $f_{\delta}$ is different for the commensurate and incommensurate cases,
\begin{equation}
f_{\delta}(u)=\left\{\begin{array}{ll}
-u\frac{\partial}{\partial u}\ln [H_{\delta}^{(2)}(u)]\qquad & \mbox{commensurate}\\
\delta\frac{\partial}{\partial\delta}\ln[I_{iu}(\delta)] & \mbox{incommensurate}
\end{array}\right.
\end{equation}
where $H_{\delta}^{(2)}$ is a Hankel function of order $\delta$ and $I_{iy}$ is a modified Bessel function with imaginary index.

\subsection{Solution using a Fokker-Planck Equation}
To demonstrate how an exact solution may be found we will derive the density of states for the commensurate case ($V(x)$ real) by using Fokker-Planck equations. Many authors have studied mathematically equivalent systems.~\cite{ov,golub,duty} Consider a general Dirac type equation,
\begin{eqnarray}
-iv_0\frac{\partial \psi_1}{\partial x}+V(x)\psi_2&=&E\psi_1\nonumber\\
iv_0\frac{\partial \psi_2}{\partial x}+V(x)\psi_1&=&E\psi_2.\label{eq:dirac}
\end{eqnarray}
The function, $V$, is a real and random function of the form,
\begin{equation}
V(x)=\Delta+\xi(x),\label{eq:def}
\end{equation}
where $\Delta$ is a constant and $\xi$ is a random field which obeys the following statistical averages,
\begin{equation}
\langle\xi(x)\rangle=0,\qquad \langle\xi(x)\xi(y)\rangle=\gamma\delta(x-y)\label{eq:average}.
\end{equation}
We reduce the Dirac equation into a system of equations for two real functions by the following transformations,~\cite{ov}
\begin{equation}
\left(\begin{array}{c}
\Psi\\
\Psi^*
\end{array}\right)=\left(\begin{array}{c}
\psi_1+\psi_2^*\\
\psi_2+\psi_1^*
\end{array}\right),\qquad\left(\begin{array}{c}
\Phi\\
-\Phi^*
\end{array}\right)=\left(\begin{array}{c}
\psi_1-\psi_2^*\\
\psi_2-\psi_1^*
\end{array}\right),
\end{equation}
then we let,
\begin{equation}
f_1=\Re\Psi,\qquad f_2=\Im\Psi,\qquad \phi_1=-\Im\Phi,\qquad \phi_2=\Re\Phi.
\end{equation}
It can be shown that $(f_1,f_2)$ and $(\phi_1,\phi_2)$ satisfy the same equations, 
\begin{equation}
\left(\begin{array}{cc}
V(x) & v_0\frac{\partial}{\partial x}\\
-v_0\frac{\partial}{\partial x} & -V(x)
\end{array}\right)\left(\begin{array}{c}
f_1\\
f_2
\end{array}\right)=E\left(\begin{array}{c}
f_1\\
f_2
\end{array}\right).\label{eq:f1f2}
\end{equation} 

Define the following function,
\begin{equation}
z=-\frac{f_2}{f_1}.
\end{equation}
By differentiating the dynamic variable, $z$, with respect to $x$ and using equation (\ref{eq:f1f2}) a dynamic equation for $z$ may be constructed,
\begin{equation}
v_0\frac{\partial z}{\partial x}=-(E-\Delta)-z^2(E+\Delta)-\xi(z^2-1).\label{eq:zd}
\end{equation}
This equation allows us to write down a Fokker-Planck equation for the random variable, $z$,
\begin{equation}
\frac{\partial P(z,x)}{\partial x}=\frac{1}{v_0}\frac{\partial}{\partial z}\left((E-\Delta)+z^2(E+\Delta)+\frac{D}{2}(z^2-1)\frac{\partial}{\partial z}(z^2-1)\right)P(z,x).
\end{equation}
The function $P(z,x)$ is the probability density distribution function of the random variable $z$ at the point $x$. It is the derivative (with respect to $x$) of the probability that $z$ is less than $x$. Since the probability of $z$ being less than infinity is unity we expect,
\begin{equation}
\int_{-\infty}^{\infty}P(z,x)dx=1.\label{eq:pdint}
\end{equation}
This is an important concept when dealing with probability densities.

We create a stationary Fokker-Planck equation by taking the limit as $x$ goes to infinity. The limit of the probability density is,
\begin{equation}
\lim_{x\rightarrow\infty}P(z,x)=p(z).
\end{equation} 
The stationary Fokker-Planck equation is,
\begin{equation}
0=\frac{1}{v_0}\frac{\partial}{\partial z}\left((E-\Delta)+z^2(E+\Delta)+\frac{D}{2}(z^2-1)\frac{\partial}{\partial z}(z^2-1)\right)p(z).
\end{equation}
This equation can be integrated and Ovchinnikov and \'Erikhman~\cite{ov} showed that the constant of integration is $N(E)$, the number of states below the energy $E$, 
\begin{equation}
N(E)=-\frac{1}{v_0}\left((E-\Delta)+z^2(E+\Delta)+\frac{D}{2}(z^2-1)\frac{\partial}{\partial z}(z^2-1)\right)p(z).\label{eq:FPl}
\end{equation}
To simplify the solution of the above differential equation we perform two transformations.

The first transformation is defined by the following function,
\begin{equation}
\cot\frac{\alpha(x)}{2}=z=-\frac{f_2}{f_1}\label{eq:alpha}.
\end{equation}
Because of the nature of the cotangent, we can obtain all possible values of the ratio, $-f_2/f_1$, by restricting $\alpha$ to some interval of length $2\pi$. We will restrict $\alpha$ to the interval $[-\frac{\pi}{2},\frac{3\pi}{2}]$. We are only interested in the case where $x\rightarrow\infty$. As in equation (\ref{eq:pdint}), the integral of all probability densities of $\alpha$ at large $x$, $p(\alpha)$, must equal unity,
\begin{equation}
\int_{-\frac{\pi}{2}}^{\frac{3\pi}{2}}p(\alpha)d\alpha=1.\label{eq:i}
\end{equation}
The relationship between $p(z)$ and $p(\alpha)$ is,
\begin{equation}
p(z)=-2p(\alpha)\sin^2\frac{\alpha}{2}.
\end{equation}
The second transformation is,
\begin{eqnarray}
\cos\alpha&=&\pm\,\mathrm{sech}\,\phi\nonumber\\
\sin\alpha&=&\pm\,\mathrm{tanh}\,\phi.
\label{eq:transf}
 \end{eqnarray}
The upper sign refers to $\alpha\in [-\frac{\pi}{2},\frac{\pi}{2}]$ and the lower sign, $\alpha\in [\frac{\pi}{2},\frac{3\pi}{2}]$. It can be shown that $p(\alpha)=p(\phi)\cosh\phi$ so that equation (\ref{eq:i}) becomes,
\begin{equation}
\int_{-\infty}^{\infty}p_{+}(\phi)d\phi+\int_{-\infty}^{\infty}p_{-}(\phi)d\phi=1,
\end{equation}  
where the $\pm$ subscript on $p$ refers to the different signs in equation (\ref{eq:transf}). After performing the two transformations equation (\ref{eq:FPl}) becomes,
\begin{equation}
\left[2E\cosh\phi\pm2\Delta+2D\frac{\partial}{\partial\phi}\right]p(\phi)=v_0N(E).\label{eq:de2}
\end{equation}
This first order differential equation can be solved with the boundary condition that $p(\phi)$ vanishes as $\phi\rightarrow\infty$, 
\begin{equation}
p(\phi)=\frac{v_0N(E)}{2D}\int_{\phi}^{\infty}dx\exp\left[\frac{E}{D}(\sinh\phi-\sinh x)\pm\frac{\Delta}{D}(\phi-x)\right].
\end{equation}

To find the number of states we recall that the probability density must be normalized so that the integral over all possible values of the random variable $\phi$ is unity. After some rearranging,
\begin{eqnarray}
2Dv_0^{-1} N(E)^{-1}=\int_{-\infty}^{\infty}d\phi\int_{\phi}^{\infty}dx\exp\left[\frac{E}{D}(\sinh\phi-\sinh x)+\frac{\Delta}{D}(\phi-x)\right]\nonumber\\
+\int_{-\infty}^{\infty}d\phi\int_{\phi}^{\infty}dx\exp\left[\frac{E}{D}(\sinh\phi-\sinh x)-\frac{\Delta}{D}(\phi-x)\right].
\end{eqnarray}
If these two integrals are combined and we let $2y=x-\phi$, $\delta=\Delta/D$ and $u=E/D$ we obtain, after changing the order of integration,
\begin{equation}
2Dv_0^{-1} N(E)^{-1}=4\int_0^{\infty}dy\cosh 2\delta y\int_{-\infty}^{\infty}d\phi\exp[u(\sinh\phi-\sinh (2y+\phi)].
\end{equation}
Now we let $z=\phi+y$,
\begin{eqnarray}
2Dv_0^{-1} N(E)^{-1}&=&4\int_0^{\infty}dy\cosh (2\delta y)\int_{-\infty}^{\infty}dz\exp[-2u\sinh y\cosh z]\nonumber\\
&=&8\int_0^{\infty}dy\cosh (2\delta y)\int_0^{\infty}dz\exp[-2u\sinh y\cosh z]\nonumber\\
&=&8\int_0^{\infty}dy\cosh (2\delta y)K_0(2u\sinh y)\nonumber\\
&=&\pi^2[J_{\delta}(u)^2+Y_{\delta}(u)^2]
\end{eqnarray}
where $J_{\delta}(u)$ is a Bessel function of index $\delta$ and $Y_{\delta}(u)$ is a Bessel function of the second kind of order $\delta$. The number of states with energy less than $E$ is,
\begin{equation}
N(E)=\frac{2D}{\pi^2v_0[J_{\delta}(\frac{E}{D})^2+Y_{\delta}(\frac{E}{D})^2]}.
\end{equation}
To find the density of states, $\rho(E)$, we differentiate $N(E)$,
\begin{equation}
\rho(E)=-\frac{4}{\pi^2v_0}\frac{J_{\delta}(u)J_{\delta}'(u)+Y_{\delta}(u)Y_{\delta}'(u)}{[J_{\delta}(u)^2+Y_{\delta}(u)^2]^2}.\label{eq:density}
\end{equation}
This is the density of states from which we obtain our results.

\section{Properties of the Commensurate solution}
\label{sec-commensurate}
\subsection{The density of states}

\begin{figure}
\centerline{\epsfxsize=8.0cm \epsfbox{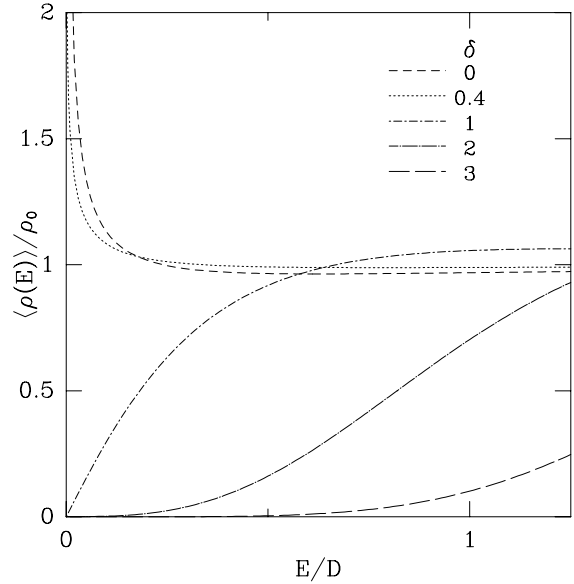}}
\vspace{0.5cm}

\caption{
Energy dependence of the disorder-averaged
density of states for the commensurate case
for various values of the dimensionless parameter
$\delta$
 (see Eq. (\protect\ref{delta})),
which is a measure of the deviation from criticality.
 The density of states is symmetrical about the Fermi energy $(E=0)$
and diverges at 
 $(E=0)$ when $\delta < {1 \over 2}.$
This parameter range corresponds to a Griffiths phase.
Note that only far from criticality
 ($\delta \gg 1$) is there effectively a gap in
the system.
  This contrasts with
the disorder-free case, for which
there is always a gap except at the critical point.
\label{figdos}}
\end{figure}

Fig. \ref{figdos} shows the energy dependence of the density of states from equation (\ref{eq:density}) for
a range of values of $\delta$ with $\rho_0=1/(\pi v_0)$.
In the low energy limit (small $u$) we can take the following approximation of equation (\ref{eq:density}),
\begin{equation}
\frac{\rho(E)}{\rho_0}=-\frac{4}{\pi}\frac{Y_{\delta}'(u)}{Y_{\delta}(u)^3}\label{eq:d}
\end{equation}
since the Bessel functions of the first kind remain finite for small $u$ whereas those of the second kind become infinite, as shown in the following small $u$ approximations,
\begin{equation}
Y_{\delta}(u)\sim\left\{\begin{array}{ll}
\frac{2}{\pi}\ln \frac{u}{2} & \delta=0\nonumber\\
-\frac{1}{\pi}\Gamma(\delta)\left(\frac{1}{2}u\right)^{-\delta}\qquad & \delta\neq 0.
\end{array}\right.\label{eq:su}
\end{equation}
These low energy limits are substituted into equation (\ref{eq:d}) and then the dominant terms are retained, that is, the smallest powers of $u$,
\begin{equation}
\frac{\rho(E)}{\rho_0}\sim\left\{\begin{array}{ll}
\frac{-\pi D}{E[\ln(\frac{E}{2D})]^3} & \delta=0\\
\frac{2\pi\delta}{\Gamma(\delta)^2}
\left(\frac{E}{2D}\right)^{2\delta-1}\qquad & \delta\neq 0.
\label{eq:y}
\end{array}\right.
\end{equation}
The divergence in the density of states at $E=0$ is
sometimes referred to as the Dyson singularity.

The function $Y_{\delta}(u)$ is continuous as $u$ and $\delta$ approach zero.
 This property is not apparent from equation (\ref{eq:y}). To avoid this problem we take the small $\delta$ (close to criticality) limit of equation (\ref{eq:d}) before we take the small $u$ limit. Before we take any limits we write the Bessel function of the second kind in terms of the Bessel function of the first kind,
\begin{equation}
Y_{\delta}(u)=\frac{J_{\delta}(u)\cos(\delta\pi)-J_{-\delta}(u)}{\sin{(\delta\pi)}}.
\label{eq:YJ}
\end{equation}
We can find a small $\delta$ approximation to $J_{\delta}(u)$ from its series expansion,
\begin{equation}
J_{\delta}(u)=\left(\frac{1}{2}u\right)^{\delta}
\sum_{k=0}^{\infty}\frac{\left(-\frac{1}{4}u^2\right)}{k\mathrm{!}\Gamma(\delta+k+1)}
\sim\left(\frac{1}{2}u\right)^{\delta}J_0(u).
\end{equation} 
So now we have a small $\delta$ approximation for a Bessel function of the second kind,
\begin{equation}
Y_{\delta}(u)=J_0(u)\frac{(\frac{1}{2}u)^{\delta}-(\frac{1}{2}u)^{-\delta}}{\delta\pi}.\label{eq:sd}
\end{equation}
When $u$ is small we set $J_0(u)=1$ and equation (\ref{eq:d}) becomes,
\begin{equation}
\frac{\rho(E)}{\rho_0}=
2\pi\delta^3\left(\frac{E}{2D}\right)^{2\delta-1}
\frac{\left[1+\right(\frac{E}{2D}\left)^{2\delta}\right]}
{\left[1-\left(\frac{E}{2D}\right)^{2\delta}\right]^3}.\label{eq:smalld}
\end{equation}
This agrees with the scaling form obtained by Balents and Fisher.~\cite{balents}
 By taking appropriate limits it can be shown that this formula
 agrees with equation (\ref{eq:y}). For $\delta=0$ we use,
\begin{equation}
\lim_{\delta\rightarrow 0}
\left(\frac{\delta}{1-\left(\frac{E}{2D}\right)^{2\delta}}\right)
=\lim_{\delta\rightarrow 0}\frac{\delta}{1-(1+2\delta\ln\frac{E}{2D})}
=-\frac{1}{2\ln\frac{E}{2D}}\label{eq:limit}
\end{equation}
and for $\delta$ becoming small we use, in equation (\ref{eq:d}),
 $\Gamma(\delta)\sim 1/\delta$.

\begin{figure}
\centerline{\epsfxsize=8.0cm \epsfbox{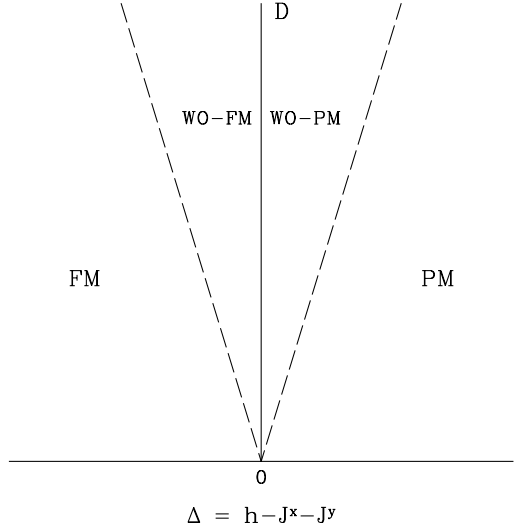}}
\caption{
Phase diagram of the Ising transition
in a random transverse field. The horizontal
axis is a measure of the deviation from criticality in the non-random
model. The vertical axis is the amount of disorder.
The four phases are: ferromagnet(FM),
 weakly ordered ferromagnet (WO-FM),
weakly ordered paramagnet (WO-PM), and paramagnet (PM).
 The weakly ordered phases are Griffiths phases
in which the linear susceptibility
diverges but there is only short-range order.
 Note that the dashed line does not represent
a true phase transition and that higher order
susceptibilities will diverge in larger regions
of the phase diagram.
\label{phased2}} \end{figure}

The   low-energy   
 ($|E|\ll D$)
dependence of 
 the density of states contains some important physics. The
density of states diverges at $E=0$ for  $\delta < 1/2$
  and is zero at $E=0$ for $\delta > 1/2 $.
These two cases lead to qualitatively very different behaviour.
In the former case some susceptibilities will
diverge as the temperature approaches zero. This
corresponds to a Griffiths or weakly-ordered phase.~\cite{griffiths}
Hence, for the Ising transition there will be four phases:
ferromagnet,
 weakly-ordered ferromagnet,
weakly-ordered paramagnet, and paramagnet~\cite{dsf}
(see Fig. \ref{phased2}).

\subsection{Ground state energy}
\label{gse}

The dependence of the ground state energy
 of the disordered commensurate system on $\delta$
 can be shown to be infinitely differentiable, but not analytic.
 To show this we follow a procedure similar to that used by McCoy and Wu~\cite{mw}
 and Shankar and Murthy~\cite{sm}
 who considered the analogous two dimensional classical system.
 To find the ground state energy in the presence of disorder we use,
\begin{equation}
\epsilon(\delta)=-\int_0^{\infty}\rho(E)EdE.
\end{equation}
We make use of the expression (\ref{eq:smalld}), 
for the density of states at low energies,
 which we assume is accurate up to an energy $E_c$, which is less than $2D$,
\begin{equation}  
\epsilon(\delta)=-\pi\rho_0 D^28\delta^3\int_0^{\frac{E_c}{2D}}\frac{E^{2\delta}(1+E^{2\delta})}{(1-E^{2\delta})^3}dE-\int_{E_c}^{\infty}\rho(E)EdE.\label{eq:E}
\end{equation}
By integrating by parts,
\begin{equation}
8\delta^3\int_0^{\frac{E_c}{2D}}
\frac{E^{2\delta}(1+E^{2\delta})}{(1-E^{2\delta})^3}dE
=\left[\frac{4\delta^2E^{1+2\delta}}
{(1-E^{2\delta})^2}\right]_0^{\frac{E_c}{2D}}-
\left[\frac{2\delta E}{(1-E^{2\delta})}\right]_0^{\frac{E_c}{2D}}
+2\delta\int_0^{\frac{E_c}{2D}}\frac{dE}{1-E^{2\delta}}.
\end{equation}
As for the disorder free case (see Section \ref{gse0})
 we subtract off the ground state energy at $\delta=0$. To calculate the $\delta=0$ ground state energy we require the limit in equation (\ref{eq:limit}), with which we obtain,
\begin{equation}
-\frac{\epsilon(0)}{\pi\rho_0 D^2}=\frac{E_c}{2D(\ln \frac{E_c}{2D})^2}+\frac{E_c}{2D\ln \frac{E_c}{2D}}-\int_0^{\frac{E_c}{2D}}\frac{dE}{\ln E}+\lim_{\delta\rightarrow 0}\frac{1}{\pi\rho_0D^2}\int_{E_c}^{\infty}\rho(E)EdE.
\end{equation}
By subtracting the zero $\delta$ case from the small $\delta$
 case and combining those terms analytic in $\delta$ in a function $f(\delta)$, we obtain,
\begin{eqnarray}
\epsilon(\delta)-\epsilon(0)&=&-\pi\rho_0 D^2\int_0^{\frac{E_c}{2D}}dE\left[\frac{2\delta}{1-E^{2\delta}}+\frac{1}{\ln E}\right]+f(\delta)\nonumber\\
&=&-\pi\rho_0 D^2\int_0^{\infty}d\xi e^{-\xi/2\delta}[(1-e^{-\xi})^{-1}-\xi^{-1}]+f(\delta)\nonumber\\
\end{eqnarray}
where we have made the substitution $E=e^{-\xi/2\delta}$ and set $E_c=2D$,
 because this does not affect the analytic properties of the integral.
 This integral can be solved in terms of Euler's $\psi$ function,~\cite{ryzhik}
\begin{eqnarray}
\epsilon(\delta)-\epsilon(0)&=&\pi\rho_0 D^2\ln 2\delta+\pi\rho_0 D^2\psi\left(\frac{1}{2\delta}\right)+f(\delta)\nonumber\\
&=&-\pi\rho_0 D^2\delta-\pi\rho_0 D^2\sum_{n=1}^{\infty}B_{2n}(2\delta)^{2n}(2n)^{-1}+f(\delta),
\end{eqnarray}
using the small $\delta$ approximation for the $\psi$ function.~\cite{ryzhik}
 The Bernoulli numbers $B_{2n}$ are proportional to
 $\frac{2n\mathrm{!}}{(2\pi)^{2n}}$ for large $n$. Because of this,
 the ground state energy has zero radius of convergence about the point $\delta=0$.
Thus the
 ground state energy is infinitely differentiable but
 is not an analytic function of $\delta$. The critical exponent, $\alpha$, defined in Table \ref{tableo} cannot be defined in this case. 

\subsection{Thermodynamic properties}

\subsubsection{Free energy}

For any
particular configuration of the disorder the free energy per site of the
system is
\begin{equation}
F = -k_BT \sum_k  \ln \left(2 \cosh \left({ E_k \over 2k_B T}
\right)\right)
\label{fe} \end{equation}
where $ \{E_k\}$ denotes the eigenvalues of the Hamiltonian
(\ref{hamel}). This simple formula holds
because the eigenstates of the Hamiltonian are non-interacting fermions.
It then follows that the disorder-averaged free energy is
\begin{equation}
\langle F \rangle  =  -k_BT \int_0^\infty
dE  \langle \rho(E) \rangle
   \ln \left(2 \cosh \left({ E \over 2k_B T}
\right)\right)
\label{feav}
\end{equation}
The low-temperature behavior of the specific
heat and the transverse susceptibility
(for the anisotropic transitions)
 follows from the energy
dependence of the disorder-averaged 
density of states.~\cite{bul,smith}
We now show this in detail.


\subsubsection{Specific heat}

The disorder averaged specific heat is,
\begin{equation}
\langle C(T)\rangle=-T\frac{\partial^2\langle F(T)\rangle}{\partial T^2}=\frac{1}{T}\int_0^{\infty}dEE^2\langle\rho(E)\rangle\frac{\partial f}{\partial E}
\end{equation}
where $f(T)$ is the Fermi distribution function.
For the commensurate case with $\delta\neq 0$ the mean specific heat is,
\begin{eqnarray}
\langle C(T)\rangle&=&\frac{2\pi\delta\rho_0}{k_BT^2\Gamma(\delta)^2(2D)^{2\delta-1}}\int_0^{E_c}dE\frac{E^{2\delta+1}e^{-E/k_BT}}{(1+e^{-E/k_BT})^2}+\frac{1}{T}\int_{E_c}^{\infty}dE\rho(E)\frac{\partial f}{\partial E}\nonumber\\
&=&\frac{2\pi\delta\rho_0 k_B^{2\delta+1}T^{2\delta}}{\Gamma(\delta)^2(2D)^{2\delta-1}}\int_0^{E_c/k_BT}dy\frac{y^{2\delta+1}e^{-y}}{(1+e^{-y})^2}+\frac{1}{T}\int_{E_c}^{\infty}dE\rho(E)\frac{\partial f}{\partial E}.
\end{eqnarray}
As the temperature becomes very small the limit of the integral, $E_c/k_BT$, becomes very large. The first integral will dominate the specific heat and will be evaluated from zero to infinity. By using integral tables~\cite{ryzhik} it can be shown that,
\begin{eqnarray}
\int_0^{\infty}dy\frac{y^xe^{-y}}{(1+e^{-y})^2}&=&\Gamma(x+1)\sum_{k=1}^{\infty}\frac{(-1)^{k+1}}{k^x},\qquad x>-1\nonumber\\
&=&\Gamma(x+1)(1-2^{1-x})\zeta(x).\label{eq:int}
\end{eqnarray}
where $\zeta$ is the Riemann zeta function. The mean specific heat is then,
\begin{equation}
\langle C(T)\rangle=\frac{4\pi\delta\rho_0k_BD\Gamma(2\delta+2)(1-2^{-2\delta})\zeta(2\delta+1)}{\Gamma(\delta)^2}\left(\frac{k_BT}{2D}\right)^{2\delta}.
\end{equation}
 For small $\delta$, $\langle C(T)\rangle\sim\delta^3T^{2\delta}$, in agreement with Fisher~\cite{dsf}
and the numerical work of Young.~\cite{young}

The specific heat in the commensurate case with $\delta=0$ is,
\begin{eqnarray}
\langle C(T)\rangle&=&-\frac{\pi D\rho_0}{k_BT^2}\int_0^{E_c}dE\frac{Ee^{-E/k_BT}}{[\ln(\frac{E}{2D})]^3(1+e^{-E/k_BT})^2}+\frac{1}{T}\int_{E_c}^{\infty}dE\rho(E)\frac{\partial f}{\partial E}\nonumber\\
&=&-\pi D\rho_0 k_B\int_0^{E_c/k_BT}dy\frac{ye^{-y}}{[\ln(\frac{k_BTy}{2D})]^3(1+e^{-y})^2}\nonumber\\
& &+\frac{1}{T}\int_{E_c}^{\infty}dE\rho(E)\frac{\partial f}{\partial E}.
\end{eqnarray}
To simplify this equation we note that the term $ye^{-y}/(1+e^{-y})^2$ is appreciable only for values of $y$ of order unity. Since we are taking a low temperature limit, $T\ll D$, we may approximate the term, $(\ln\frac{yk_BT}{D})^3$, to simply $(\ln\frac{k_BT}{D})^3$ when $y\sim 1$. As in the previous case, the second integral is negligible as the temperature approaches zero and the limits of the first integral are zero and infinity. The specific heat for $\delta=0$ is
\begin{equation}
\langle C(T)\rangle=-\frac{\pi k_BD\rho_0}{(\ln\frac{k_BT}{2D})^3}\ln 2,
\end{equation}
which has the same temperature dependence as found by Fisher.~\cite{dsf}

\subsubsection{Transverse susceptibility}

The mean transverse field susceptibility is, for the anisotropic transition,~\cite{smith}
\begin{equation}
\langle\chi_{zz}(T)\rangle=\int_0^{\infty}dE\langle\rho(E)\rangle\frac{\partial f}{\partial E}.
\end{equation}
For the commensurate case with $\delta\neq 0$ and the temperature approaching zero the calculation of the susceptibility is similar to the calculation of the specific heat with $\delta\neq 0$. The mean transverse susceptibility is,
\begin{equation}
\langle\chi_{zz}(T)\rangle=\frac{2\pi\delta\rho_0\Gamma(2\delta)(1-2^{2-2\delta})\zeta(2\delta-1)}{\Gamma(\delta)^2}\left(\frac{k_BT}{2D}\right)^{2\delta-1}.
\end{equation}
If $\delta<\frac{1}{2}$ and $T\rightarrow 0$ the susceptibility becomes infinite. This is the Griffiths phase region. We cannot take the limit as $\delta$ goes to zero of the susceptibility since the condition on the integral in equation (\ref{eq:int}) is $x>-1$ which means, in this case, $\delta>0$. If we try to take this limit we see that it does not exist. The critical exponent, $\gamma$, is not defined.

Similarly to the specific heat calculation with $\delta=0$ it can be shown that when $\delta=0$ the mean susceptibility at low temperatures is,
\begin{equation}
\langle\chi_{zz}(T)\rangle=-\frac{\pi\rho_0D}{k_BT}\int_0^{\infty}dy\frac{e^{-y}}{y(\ln[\frac{yk_BT}{2D}])^3(1+e^{-y})^2}.
\end{equation}
The integrand is large at $y=1$ and $y=0$ so we can approximate the integral to an integral from zero to some cut off, $A$, of order unity. We also notice that, for $y\ll 1$,
\begin{equation}
\frac{e^{-y}}{y(\ln[\frac{yk_BT}{2D}])^3(1+e^{-y})^2}\sim \frac{1}{y(\ln[\frac{yk_BT}{2D}])^3}.
\end{equation}
Using these approximations the susceptibility is,
\begin{eqnarray}
\langle\chi_{zz}(T)\rangle&=&-\frac{\pi\rho_0D}{k_BT}\int_0^Ady\frac{1}{y(\ln[\frac{yk_BT}{2D}])^3}\nonumber\\
&=&\frac{\pi D\rho_0}{k_BT}\left[\frac{1}{[\ln(\frac{yk_BT}{2D})]^2}\right]_0^A\nonumber\\
&=&\frac{\pi D\rho_0}{k_BT[\ln(\frac{Ak_BT}{2D})]^2}.
\end{eqnarray}
This susceptibility is finite when $\delta=0$ unless $T=0$. A comparison with the $\delta\neq 0$ result shows that the susceptibility is not continuous at the phase transition which is at $\delta=0$.

\subsection{Dynamic critical exponent $z$}

This relates the scaling of energy (or time) scales
to length scales.
We can make the following crude scaling argument
to extract $z$ from the low-energy behavior of the 
density of states.
The total number of states (per unit length) with energy less than
$E$, $N(E)$ scales with the inverse 
of any length scale $\ell$. By definition $E \sim \ell^{-z}$.
This implies that
$   \langle \rho(E) \rangle  \sim E^{1/z -1}.$
Thus for the commensurate case, to leading order in $\delta$,
\begin{equation}
z= {1  \over  2 \delta },
\label{zz}
\end{equation}
in agreement with the renormalization group
results of Fisher~\cite{dsf} and the numerical results
of Young and Rieger.~\cite{yr}
This is a particularly striking result because it shows
that (i)    $z$ is not universal and
(ii)    $z$     diverges at the critical point.
The latter implies logarithmic scaling and
activated dynamics.~\cite{ry}

\subsection{Finite size scaling}

Monthus {\it et al.}~\cite{mon} studied          
an equation equivalent to (\ref{hamel})
with $V(x)$ real and $\Delta=0$.~\cite{bouch}
They have shown that  on a line of length $L$,
for a typical potential $V(x)$
the lowest eigenvalue $E_0$ scales like
$E_0^2 \sim \exp(-c L^{1/2})$, where $c$
is a constant.
This is consistent with the scaling of
$\ln E_0$ with $L^{1/2}$ at the critical point
found numerically.~\cite{yr}
The average $\langle E_0^2 \rangle \sim \exp(-d L^{1/3})$ 
where $d$ is a constant,~\cite{mon}
showing the discrepancy between {\it average} and
{\it typical} values.
 Fisher and Young~\cite{dsf3}
recently derived the distribution function for the energy gap
from the RSRGDT and compared it to numerical results.
The distribution function they derived gives average and
typical values in agreement with the above results.

\subsection{Correlation lengths}

 Fisher~\cite{dsf} stressed the
distinction between average and
 typical correlations.
If $C_{ij}\equiv \langle A_i   A_j \rangle$
denotes a correlation function of a variable $A_i$
then the average correlation function
$C_{av}(r) \equiv {1 \over L} \sum_{i=1}^L C_{i,i+r}$
is what is measured experimentally.
Away from the critical point $C_{av}(r) \sim \exp(-r/\xi_{av})$
where $\xi_{av}$ is the average
correlation length.
However, $C_{av}(r)$ is dominated by rare 
pairs of spins with $C_{ij} \sim 1.$
In contrast, with probability one $C_{i,i+r}
 \sim \exp(-r/\xi_{typ})$
where $\xi_{typ}$ denotes the typical
correlation length. It  is distinctly different
from $\xi_{av}$ ($\xi_{typ} \ll \xi_{av} $),
 having  a different critical exponent.
The localization length is useful because it is proportional
to the typical correlation
length for quantities that are
diagonal in the fermion representation.~\cite{klein}
 
The localization length is obtained from integrating equation (\ref{gf}),
\begin{equation}
\frac{1}{\lambda(E)}=\frac{D}{v_0}\Re(f_{\delta}(u))+\mbox{constant}.
\end{equation}
In the commensurate case the following approximation holds for small $u$,
\begin{equation}
\Re(f_{\delta}(u))=-\frac{uY_{\delta}'(u)}{Y_{\delta}(u)}.
\end{equation}
Equations (\ref{eq:su}) and (\ref{eq:sd}) give, for small energy,
\begin{equation}
\lambda(E)=\left\{\begin{array}{ll}
\frac{v_0}{D\delta} & \delta\neq 0\\
-\frac{v_0}{D}\ln\frac{E}{2D} & \delta=0.
\end{array}\right.
\end{equation}
The localization length is infinite only when $\delta=0$ and $E=0$.

For both the pure system and the random system,
 $\xi_{typ}\sim\lambda(0)^{-1}\sim\Delta^{-1}$,
 indicating that $\nu_{typ}=1$ and that
this critical exponent is not modified by the presence of disorder.
This result also agrees with the RSRGDT.
Balents and Fisher studied the same Dirac equation
and examined the decay of the average Green function.
Hence, they found the critical exponent associated with the average
 correlation length, $\xi_{av}\sim\Delta^{-2}$.

\section{Properties of the incommensurate solution}
\label{sec-incommensurate}

It was shown in section \ref{sec-caseII} that the incommensurate solution
 with $\delta=\frac{\Delta}{D}=0$
 describes the $XX$ chain which has no anisotropy.
 We can use equation (\ref{gf}) to find the density of states. Alternatively, using Fokker-Planck equations, or a number of other methods,~\cite{hayn,golub,gor,abr} it can be shown that the number of states below $E$ is,
\begin{equation}
N(E)=\frac{D\rho_0}{\pi}\frac{\sinh(\pi u)}{|I_{iu}(\delta)|^2},
\end{equation}
where $u=E/D$. The density of states, $\rho$, may be found by taking the derivative of $N(E)$. The density of states for small $u$ is, 
\begin{equation}
\frac{\rho(E)}{\rho_0}=\frac{1}{I_0(\delta)^2}.
\end{equation}
When $\delta=0$, for any $u$,
\begin{equation}
\rho(E)=\rho_0.
\end{equation}
Hence, the incommensurate density of states is always finite.
 When $\delta=0$ the density of states is constant.
Thus for an XX random chain in a non-zero transverse field
there is no Dyson singularity. This agrees with the
results of Smith.\cite{smith}

To find the localization length we take the real part of equation (\ref{gf}) and integrate over $E$. For small $u$, the localization length for the incommensurate case is,
\begin{equation}
\lambda(E)=\frac{4v_0}{D}\left[1+\frac{4\delta I_1(\delta)}{I_0(\delta)}\right]^{-1}.
\end{equation}
The constant of integration must
be evaluated by deriving the localization length
 from other methods.~\cite{gogolin} The 
most important property of this result is that
unlike for the commensurate case  
the localization length is always finite.
This means that the typical correlation length of
the corresponding spin model does not
diverge when the pure system is at criticality.
Hence, in a non-zero transverse field the anisotropy
phase transition does not occur if there is randomness
in the transverse field or the isotropic exchange.

\section{Conclusions}

We presented some exact results for the effect
of disorder on the quantum critical properties
one of the simplest models to undergo
quantum phase transitions:
an anisotropic XY spin chain in a transverse
field. By taking the continuum limit of the corresponding
non-interacting fermion model we were able to map 
various cases of the model onto a
a Dirac equation with a random mass.
This mapping has the distinct advantage that 
a number of different techniques can then be used to
obtain exact analytic results for the density of states and the localization
length. In the presence of disorder 
Ising transition of the model is in the same universality
class as the random transverse field Ising model.
If there is only randomness in the anisotropy 
then the anisotropy transition is also in this universality class.
However, if there is randomness in the isotropic part of
the exchange or in the transverse field then
in a non-zero transverse field the anisotropy transition
is destroyed by the disorder.
By examining the energy dependence of the density
of states we showed that the dynamical critical exponent,
show the existence of a  Griffiths' phase near the transition, and show
that the ground state energy has an
essential singularity at the transition. 
The results obtained for the typical correlation length, the dynamical critical exponent,
the finite-size scaling of the energy gap,
and for the temperature dependence of the specific heat
near the Ising transition agree with the
results of the RSRGDT and numerical work.
Since our result is explicitly exact
this agreement is consistent with Fisher's claim that
the RSRGDT gives exact results for critical behaviour.
The real challenge is whether the mapping to the
fermion model used here can be used to obtain
results for distribution functions and spin correlation functions.
Recently some has been done on
distribution functions associated with the
zero energy eigenstates of the random Dirac equation.\cite{balents,shelton,steiner,steiner2}

\acknowledgements

This work was supported by the Australian Research Council.
We thank  R. J. Bursill, V. Dobrosavljevic, D. S. Fisher,
D. Huse, R. Hyman, V. Kotov, S. Sachdev, T. Senthil,
and R. Shankar for helpful discussions.
D. Scarratt helped produce some of the figures.

\newpage
\begin{table}
\caption{
 Experimental realizations of random spin chains.
These chains all involve antiferromagnetic exchange
except for Sr$_3$CuPt$_{1-x}$Ir$_x$O$_6$
which involves both ferromagnetic and
antiferromagnetic exchange.}
\begin{tabular}{lcc}
Material & Spin per site  & Reference \\
 \tableline
Quinolinium(TCNQ)$_2$ & ${1 \over 2}$
 & \protect\onlinecite{bul} \\
Sr$_3$CuPt$_{1-x}$Ir$_x$O$_6$ & ${1 \over 2}$ &
\protect\onlinecite{nguyen}\\
MnTPP-TCNE(solvent)& alternating ${1 \over 2}$ and $2$
 & \protect\onlinecite{epstein}\\
MgTiOBO$_3$ (warwickite) &   ${1 \over 2}$ 
 & \protect\onlinecite{warwick}\\
MgVOBO$_3$ (warwickite) &  1  &  \protect\onlinecite{warwick2}\\
Cu(3-methylpyridine)$_2$Cl$_2$& ${1 \over 2}$ &  
\protect\onlinecite{wolt} \\
\end{tabular}
\label{table0}
\end{table}

\begin{table}
\caption{
Critical exponents for the transition
in the transverse field Ising chain, without and with disorder.
$\Delta$ is a measure of the deviation from the critical point.
The exponents for  the random case were calculated
by Fisher.~\protect\cite{dsf} Some exponents
are expressed in terms of the golden mean,
$\phi \equiv {1 \over 2} ( 1 + \surd{5})$.
$H$ is an external field in the $x$ direction,
i.e., the same direction as the magnetization.
The exponents $\alpha$ and $\gamma$ are not defined (n.d.)
in the random model due to the presence of the Griffiths phase.
With disorder $\delta = \infty$
because $\langle \sigma^x_i \rangle \sim 
(\ln ( 1/|H|))^{-\beta}$.
}
\begin{tabular}{lccc} Exponent & Definition  & No  disorder & with disorder \\ \tableline
$\alpha$& $\epsilon  \sim \Delta^{2 - \alpha}$ & $0^+$(log) & n.d. \\
$\beta$ & $\langle \sigma^x_n \rangle \sim \Delta^\beta$
 & 1/8 & $2 - \phi $\\
$\gamma$ & $\chi_{xx} \sim \Delta^\gamma$ & 7/4 & n.d. \\
$\delta$ & $\langle \sigma^x_n \rangle \sim H^{1/\delta} $ ($\Delta=0$)
& 15 & $\infty$ \\
$\nu$ & $\xi_{av} \sim \Delta^{-\nu}$ & 1 & 2 \\
$\eta$ & $\langle \sigma^x_r \sigma^x_0 \rangle \sim r^{1-\eta}$
($\Delta=0$)
& 5/4 & $\phi - 1 $ \\
 $z$ & $\tau \sim \xi^z$ &1  & $\infty$ \\
\end{tabular}
\label{tableo}
\end{table}

\newpage

\newpage

\end{document}